\documentclass[%
 reprint,
superscriptaddress,
 amsmath,amssymb,
longbibliography,
]{revtex4-1}

\usepackage{array}
\usepackage{calc}

\usepackage{etoolbox,siunitx}
  \robustify\bfseries

  \usepackage{grffile}
  \usepackage{graphicx}
  \usepackage{color}
  \usepackage{dcolumn}
  \usepackage{bm}
\usepackage{multirow}
\usepackage{pbox}
\usepackage{tabularx,booktabs} 

  \newcommand{\angstrom}{\textit{$\text{\AA}$}}

\definecolor{pink}{rgb}{0.858, 0.188, 0.478}
\definecolor{green}{rgb}{0, 0.4, 0}

\makeatletter
\def\NAT@def@citea{\def\@citea{\NAT@separator}}
\makeatother

\begin{document}
  \title{Control of interlayer physics in 2H transition metal dichalcogenides }


\author{Kuang-Chung Wang}
\affiliation{School of Electrical and Computer Engineering, Purdue University, West Lafayette, IN 47906, USA}
\author{Teodor K. Stanev}
\affiliation{Department of Physics and Astronomy, Northwestern University, Evanston, Illinois 60208, USA}
\author{Daniel Valencia}%
\affiliation{School of Electrical and Computer Engineering, Purdue University, West Lafayette, IN 47906, USA}
\author{James Charles}
\affiliation{School of Electrical and Computer Engineering, Purdue University, West Lafayette, IN 47906, USA}
\author{Alex Henning}
\affiliation{Department of Materials Science and Engineering, Northwestern University, Evanston, Illinois 60208, USA}
\author{Vinod K. Sangwan}
\affiliation{Department of Materials Science and Engineering, Northwestern University, Evanston, Illinois 60208, USA}
\author{Aritra Lahiri}
\affiliation{School of Electrical and Computer Engineering, Purdue University, West Lafayette, IN 47906, USA}
\affiliation{Department of Electrical Engineering, IIT Bombay, India}
\author{Daniel Mejia}
\author{Prasad Sarangapani}
\affiliation{School of Electrical and Computer Engineering, Purdue University, West Lafayette, IN 47906, USA}
\author{Michael Povolotskyi}
\affiliation{Network for Computational Nanotechnology, Purdue University, West Lafayette, IN 47906, USA}
\author{Aryan Afzalian}
\affiliation{TSMC, Kapeldreef 75, 3001 Leuven, Belgium}
\author{Jesse Maassen}
\affiliation{Department of Physics and Atmospheric Science, Dalhousie University, Halifax, Nova Scotia, Canada, B3H 4R2.}
\author{Gerhard Klimeck}
\affiliation{School of Electrical and Computer Engineering, Purdue University, West Lafayette, IN 47906, USA}
\affiliation{Network for Computational Nanotechnology, Purdue University, West Lafayette, IN 47906, USA}
\affiliation{Purdue Center for Predictive Materials and Devices, Purdue University,  West Lafayette, IN 47906, USA}
\author{Mark C. Hersam}
\affiliation{Department of Materials Science and Engineering, Northwestern University, Evanston, Illinois 60208, USA}
\affiliation{Department of Chemistry, Northwestern University, Evanston, Illinois 60208, USA}
\author{Lincoln J. Lauhon}
\affiliation{Department of Materials Science and Engineering, Northwestern University, Evanston, Illinois 60208, USA}
\author{Nathaniel P. Stern}
\affiliation{Department of Physics and Astronomy, Northwestern University, Evanston, Illinois 60208, USA}
\author{Tillmann Kubis}
\affiliation{School of Electrical and Computer Engineering, Purdue University, West Lafayette, IN 47906, USA}
\affiliation{Network for Computational Nanotechnology, Purdue University, West Lafayette, IN 47906, USA}
\affiliation{Purdue Center for Predictive Materials and Devices, Purdue University,  West Lafayette, IN 47906, USA}

 \email{Second.Author@institution.edu}

\date{\today}

  \begin{abstract}
  
    It is assessed in detail both experimentally and theoretically how the interlayer coupling of transition metal dichalcogenides controls the electronic properties of the respective devices. Gated transition metal dichalcogenide structures show electrons and holes to either localize in individual monolayers, or delocalize beyond multiple layers - depending on the balance between spin-orbit interaction and interlayer hopping. This balance depends on layer thickness, momentum space symmetry points and applied gate fields. The design range of this balance, the effective Fermi levels and all relevant effective masses is analyzed in great detail. A good quantitative agreement of predictions and measurements of the quantum confined Stark effect in gated MoS$_2$ systems unveils intralayer excitons as major source for the observed photoluminesence.
  \end{abstract}

\maketitle

\section{Introduction}
Transition metal dichalcogenides (TMDs) are expected to push nanotechnology to the ultimate scaling limit of one or a few atoms only. In contrast to graphene, these 2D materials maintain a native bandgap that is essential for most electronic device applications. TMD based devices have excellent sensitivity to external fields~\cite{Novoselov2005,Radisavljevic2011,Yun2012b}. 
Obvious ultrascaled applications range from sensing (e.g. Refs.~\cite{Zhu2017,Jariwala2014,Kalantar-zadeh2016,Esmaeili-Rad2013}),
lighting (e.g. Refs.~\cite{Ross2014,Choi2014,Withers2015}), logic devices (e.g. Refs.~\cite{Bertolazzi2013,Sangwan2015}) and wearable electronics (e.g. Refs.~\cite{Zheng2016,Oh2016}). 

The weak van der Waals coupling between TMD layers
allows for low cost fabrication (with
micro-mechanical cleavage~\cite{Bonaccorso2012}) and stacking of different TMD materials on top of each other which significantly widens the material design space~\cite{Britnell2013,Geim2014}.
A critical question for most TMD systems is the nature of the coupling between TMD layers. 
Properties of TMD materials can be tuned via the number of coupled layers.
For instance, the band-gap of MoS$_{2}$ varies by about $1$~eV
and switches between direct and indirect when the thickness changes~\cite{Yun2012b}. 
These features suggested to combine
TMD layers of different thicknesses and materials to improve (e.g.) tunneling field effect transistors~\cite{Howell2015b,chen2016thickness}. 

In spite of the importance of the interlayer coupling, its detailed properties and dependencies are not fully assessed, yet. For instance, recent experiments on excitons in TMD materials raised the question whether
electron hole recombinations are predominantly within the same or between
different TMD layers~\cite{Zhu2015,Chen2016}. 
Depending on the experimental setup, interlayer excitons appear in photoluminesence measurements, while intralayer transitions yield either a finite or a vanishing Stark effect~\cite{Chu2015,Klein2016}. 
Given the varying findings in literature, a comparative study of experiments and realistic theoretical models is needed to conclusively assess the interlayer coupling. This is the core purpose of this work. 

Although the main focus of this work is the theoretical assessment of the interlayer physics, the reliability of the theoretical answers is assessed with quantitative comparisons of predicted quantum confined Stark effects with experimental observations in various gated MoS$_2$ structures. 

All TMD devices in this work are subatomically resolved. Ab-initio electronic Hamiltonian operators are discretized with maximally localized Wannier functions (MLWF)~\cite{Marzari2012a}. This treatment combines numerical efficiency with the best known physical accuracy~\cite{Szabo2015}. 
In contrast to pure ab-initio models~\cite{Chu2015}, this approach allows to realistically include the presence of electric gates, thickness dependent doping and dielectric constants. 
Charge effects turn out to significantly influence the interlayer coupling. Commonly, electronic charge distributions are interpreted point-like within the discretization of the Poisson equation~\cite{Ilatikhameneh}. It is an important aspect of this work that these charges are resolved in subatomic resolution as well. These features guarantee full transferability of the electronic model~\cite{Szabo2015,Fang2015} to device dimensions that are computationally inaccessible to pure density functional theory (DFT) applications~\cite{Maassen2013}.
Note that important device aspects such as doping densities and spatially varying gate control are beyond the scope of pure DFT applications otherwise~\cite{Kresse2009}.

All fabricated devices of this work consist of varying numbers of TMD layers
placed on SiO$_{2}$ and highly p-doped Si that serves as gate electrode. 
Our calculations show both interlayer and intralayer
excitons yield Stark effects. However, intralayer transitions are about 2
orders of magnitude less likely. Our experimental data and theoretical predictions for the Stark effect of intralayer transitions agree well. 
Both theory and experiments do not show significant interlayer transitions.

  \begin{figure}[!htbp]
  \centering
  \includegraphics[width=2.5in]{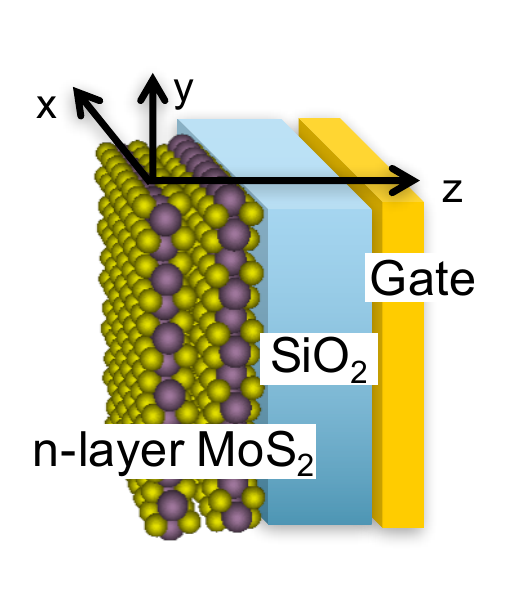} 
  \caption{Schematics of the metal-oxide semiconductor structure considered in this paper. The MoS$_2$ structure varies in thickness and is limited to the right by the gate and to the left by vacuum.}
  \label{fig:structure}
  \end{figure}

The paper is organized as follows. In the next section, details of the theoretical model and 
experimental setup are given.
The results section first confirms the transferability of the electronic bandstructure model to the considered devices. 
It then illustrates and discusses wave function and band structure properties and their dependence on the balance of spin-orbit interaction and interlayer coupling strength. This balance depends on layer thickness, electronic momentum and applied gate fields. 
The comparison of calculated band gap changes with experimentally observed photoluminesence data confirms Stark effects for both inter and intra layer direct band gap excitons. In agreement with literature, intralayer excitons are found to be significantly more visible.

\section{Method}
\subsection{Experiment}
The device fabrication followed the procedure of Ref.~\cite{Bonaccorso2012}.
MoS$_2$ samples were mechanically exfoliated from bulk MoS$_2$ crystals and put on \SI{285}{\nano\meter} thick SiO$_2$ layers that were grown on p-doped silicon substrates (see schematic in Fig.~\ref{fig:structure}). Suitable multilayer samples were identified by optical microscopy and their thickness was confirmed independently through atomic-force microscopy.
All fabricated samples were first annealed at 300 degrees Celsius to remove residue from the exfoliation process in Ar/H2 environment (800/200 SCCM). The electrostatic gates were defined through e-beam lithography on a MMA/PMMA mask, followed by 100nm deposition of Au via thermal evaporation. 
Next, the devices were annealed at 200 degrees Celsius (800/200 SCCM, Ar/H2) to clean the substrate of polymer residue from fabrication and improve the contact between metals and semiconductor layers before being loaded into a closed-cycle optical cryostat (Advanced Research Systems DMX-20-OM)~\cite{manual_ARS} and pumped down to high vacuum at 10$^{-7}$ mbar. 
The devices were kept at \SI{350}{\kelvin} for about 6 hours in high vacuum while laser annealing was applied with a \SI{5}{\milli\watt} beam of a \SI{100}{\micro\meter} spot size. 
This was meant to help remove atmospheric contaminants, before the samples were cooled down to 10 K, i.e. the temperature at which all measurements were conducted.
Photoluminescence measurements were performed using a continuous pump laser beam of 532 nm wavelength with a laser power of \SI{40}{\micro\watt} and a spot size of \SI{2}{\micro \meter}. The emitted light was analyzed with an Andor Shamrock Spectrograph~\cite{manual_spec} using a 150 lines/mm grating.

\renewcommand{\arraystretch}{1.5}
\begin{table}[h!]
\centering
 \begin{tabular}{||c c c c c||} 
 \hline
 & u (\angstrom) & a(\angstrom) & c(\angstrom)& range(\angstrom)\\ 
 \hline\hline    
 MoS$_2$ & 3.12 & 3.18 & 12.48 &  20 \\ 
 MoSe$_2$& 3.34 & 3.32 & 13.14 &  25 \\
 WS$_2$  & 3.15 & 3.19 & 12.49 &  23 \\
 MoTe$_2$& 3.62 & 3.56 & 14.22 &  26 \\
 WSe$_2$ & 3.36 & 3.33 & 13.24 &  26 \\ 
 \hline
 \end{tabular}
   \caption{\label{tb:mater_param} Structure parameters of all TMD materials resulting from the relaxation and parameterization algorithm described in the main text.}
\end{table}

\subsection{Model}
\begin{figure}[!htbp]
  \centering
  \includegraphics[width=3in]{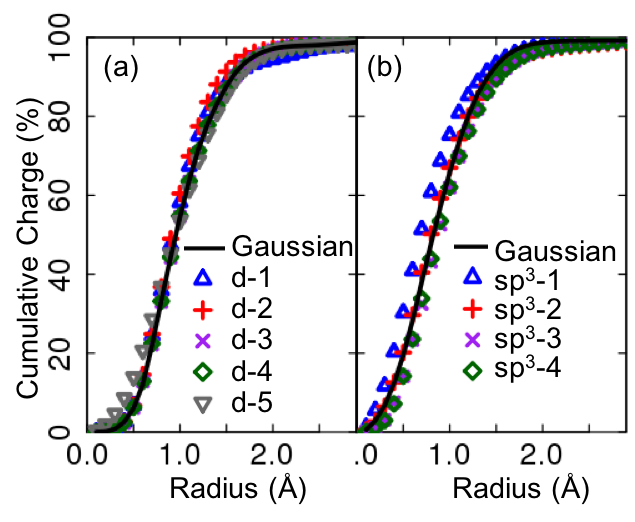}
  \caption{ Integrated charge contribution function $P(r_0)$ as defined in Eq.~\ref{eqn:cumulative} as a function of the integration radius for orbitals of molybdenum (a) and sulfur (b) atoms in infinitely thick MoS$_2$. The black lines show $P(r_0)$ when the orbital wave functions are approximated with Gaussian functions of $\sigma=0.68\angstrom$ (a) and $\sigma=0.6\angstrom$ (b), respectively.}
  \label{fig:charge_cumulative}
  \end{figure}

The atomic structures of all TMD layers modelled in NEMO5~\cite{Steiger2011} (i.e. MoS$_2$, WTe$_2$, WS$_2$, WSe$_2$, and MoSe$_2$) are based on  relaxation calculations of the respective infinite number of layers system in trigonal prismatic polytype (i.e., in 2H symmetry)~\cite{Yun2012b} performed in the DFT tool VASP~\cite{Kresse1996} with the self-consistent electronic model and the convergence criterion of \SI{1e-8}{\eV}. 
A momentum mesh of 5$\times$5$\times$5 Monkhorst-Pack grids and energy cutoff of \SI{520}{\eV} is used along with van der Waals force included according to Ref.~\cite{doi:10.1021/jp106469x}. 
The lattice constants deduced from these DFT based relaxation calculations are given in Table~\ref{tb:mater_param} and agree well with the findings in Ref.~\cite{JELLINEK1960,Wei2015}.  
The applied DFT model is based on the generalized gradient approximation utilizing the Perdew-Burke-Ernzerhof  functionals~\cite{Perdew1996}. 
The electronic DFT Hamiltonian is transformed into an MLWF representation using the Wannier90 software~\cite{Marzari1997,PhysRevB.72.125119,Das2017a} with $d$ orbitals for the metal electrons and $sp^{3}$ orbitals for the chalcogenide electrons as the initial projection. 
The spreading of the Wannier functions~\cite{Marzari2012a} is reduced iteratively until it converges to around 2~$\angstrom^2$. Atom positions and their corresponding electronic Hamiltonian of finite TMD structures are then created in NEMO5~\cite{Steiger2011} as portions of the respective infinite system.
As a consequence, all TMD systems in this work are intralayer periodic (in x- and y-direction of the schematic in Fig.~\ref{fig:structure}) with Bloch boundary conditions applied. 
Nonlocal Hamiltonian elements are considered up to the material specific range listed in  Table~\ref{tb:mater_param}.
All calculations of gated MoS$_2$ structures sketched in Fig.~\ref{fig:structure} are performed with self-consistent solutions of the Schr\"{o}dinger and Poisson equations. The Poisson equation is discretized on a finite element mesh (FEM). The resulting electrostatic potentials converged for FEM resolutions of  $0.6\angstrom$ or better. The electronic density resulting of the solution of the Schr\"{o}dinger equation is transformed into real space simplifying the MLWF basis with  Gaussian functions with $\sigma$=0.68 $\angstrom$ (in Eq.~\ref{eqn:gaussian}). This simplification eases the numerical burden during the iterative solution of the Poisson and the Schr\"{o}dinger equation significantly and does not noticeably alter the actual spatial charge distribution as illustrated in Fig.~\ref{fig:charge_cumulative}. 
This figure shows the integrated charge contribution function $P(r_0)$ in Eq.~\ref{eqn:cumulative} solved with the charge distribution function $\rho(r,\theta,\phi)$ of the MLWF and the fitted Gaussian function $\rho_{G}(r,\theta,\phi)$ of Eq.~\ref{eqn:gaussian}, respectively.

\begin{equation}
\label{eqn:gaussian}
\rho_{G}(r,\theta,\phi)= \frac{1}{\sqrt{8\sigma^{6}\pi^3}}  \exp\left(-\frac{r^2}{\sigma^2}\right)
\end{equation}

\begin{equation}
\label{eqn:cumulative}
P(r_0)= \frac{ \int_0^{\pi} \int_0^{2\pi} \int_0^{r_0} \rho (r,\theta,\phi) r^2 sin \phi \, dr d\theta d\phi }{ \int_0^{\pi} \int_0^{2\pi} \int_0^{\infty} \rho (r,\theta,\phi) r^2 sin \phi \, dr d\theta d\phi   } 
\end{equation}

The MoS$_2$ thickness of the gated structure in Fig.~\ref{fig:structure} is varied between one and ten layers.
In these cases, the donor doping is set to \SI{1.5 e18}{\per\cubic\centi\metre} for MoS$_2$ monolayers, \SI{2e19}{\per\cubic\centi\metre} for 6 MoS$_2$ layers. 
The effective doping is induced from the atmospheric adsorbates. Their values are deduced from experimental threshold voltages and gate oxide capacitances~\cite{Jariwala2013b}.
The doping density is linearly interpolated for MoS$_2$ layer systems in between 1 and 6 layers and it is assumed to be saturated for MoS$_2$ layers thicker or equal to 6-layers~\cite{Jariwala2013b}.
As commonly done in device calculations~\cite{1004231}, the computational burden of \SI{285}{\nano\meter} thick SiO$_2$ as gate dielectric is avoided with a 12nm dielectric slab of the same equivalent capacitance in the Poisson equation. SiO$_2$ did not enter the electron density calculations.
The gate is considered as a Schottky contact with a metal work function of \SI{5.15}{\eV}\cite{Novikov2010} for the highly p-doped Si. 
The energy offset is set to the Fermi level of the device electrons. 
The uncapped TMD side is considered to be exposed to vacuum, modeled with vanishing field boundary conditions for the Poisson equation. 
The MoS$_2$ dielectric constant is assumed to be homogeneous but linearly varying with the layer thickness following Ref.~\cite{Chen2015a}. 
The electronic wave functions of the conduction and the valence band states are used to solve the optical transition matrix elements. 
Peaks in these optical elements are considered as optical transition energies~\cite{7301982}. 
For all gate-independent bandstructures and wave function assessments in this work, a constant doping of \SI{1.5 e18}{\per\cubic\centi\metre} is assumed and the Fermi level is chosen to achieve local charge neutrality for the respective systems. The electron hole recombination energies are extracted from  single particle bandstructures. As discussed in Ref.~\cite{Liu1}, many particle effects are expected to have in low order no net impact on the transistion energies.

  \begin{figure}[!htbp]
  \centering
  \includegraphics[width=3.4in]{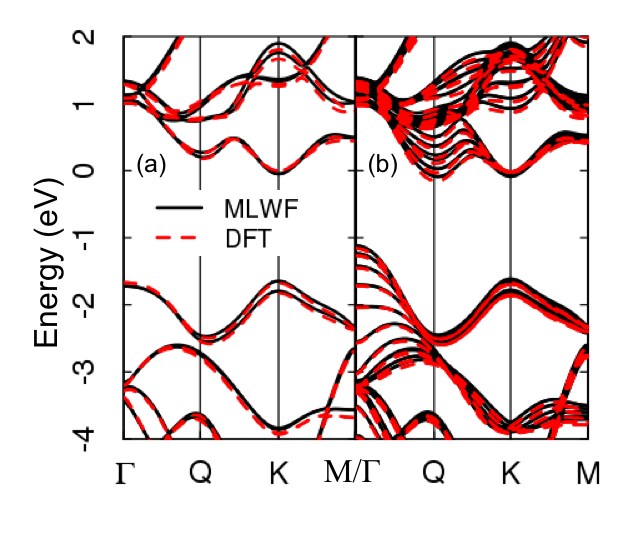}
  \caption{ Comparison of mono-layer (a) and quintuple layer (b) MoS$_2$ band diagrams solved with MLWF in NEMO5 and the DFT functionality of VASP. The agreement confirms the transferability of the MLWF parameters. }
  \label{fig:EK1and5layer}
  \end{figure}

  \begin{figure}[!htbp]
  \centering
  \includegraphics[width=3.4in]{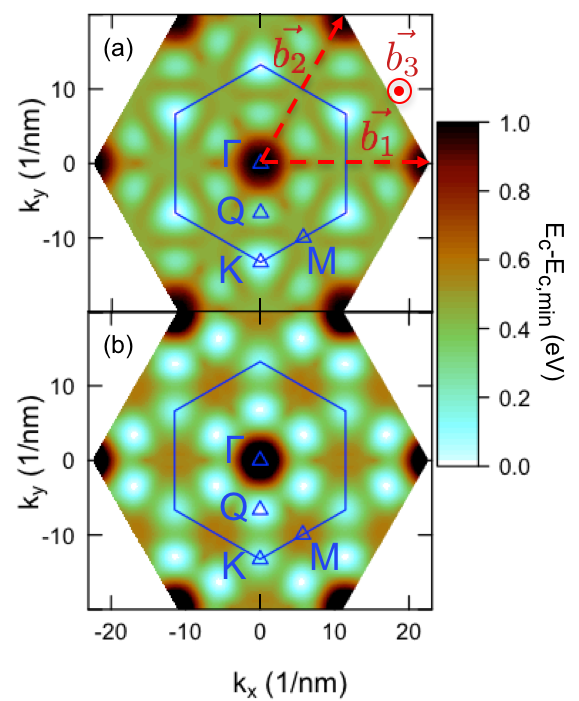}
  \caption{Contour graph of the conduction band minimum of mono-layer (a) and 5-layer (b)  MoS$_2$. The blue hexagon depicts the first Brillouin zone. Reciprocal lattice vectors are labeled with $\vec{b_1}$, $\vec{b_2}$ and $\vec{b_3}$. Note that the location of the Q valley is close to the middle  between K and $\Gamma$.  }
  \label{fig:EK1and5layercontour}
  \end{figure}

  \section{Results}

  \subsection{Transferability of MLWF parameters}
  
The NEMO5-calculated bandstructures in the MLWF representation agree well with the ab-initio results of the VASP software~\cite{Yun2012b} for any MoS$_2$ layer thickness (see Fig.~\ref{fig:EK1and5layer} for the monolayer and 5-layer cases). Very similar transferability of the MLWF representation and fitting procedure was found for all other TMD materials and layer thicknesses. 
This is remarkable, since MLWF parameterizations are sometimes created for each material thickness individually~\cite{Szabo2015}.

  \begin{figure}[!htbp]
  \centering
  \includegraphics[width=3.4in]{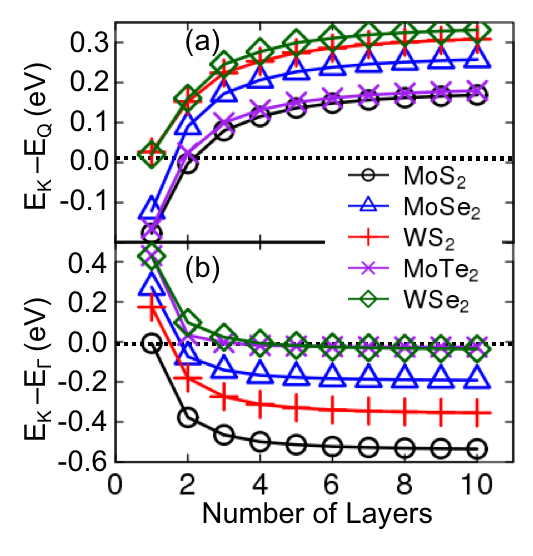}
  \caption{Valley energy differences in the  conduction band  (K and Q valley) (a) and valence band (K and $\Gamma$ valley) (b) as a function of the TMD layer thickness.}
  \label{fig:valley}
  \end{figure}
  
    \begin{figure}[!htbp]
  \centering
  \includegraphics[width=3.4in]{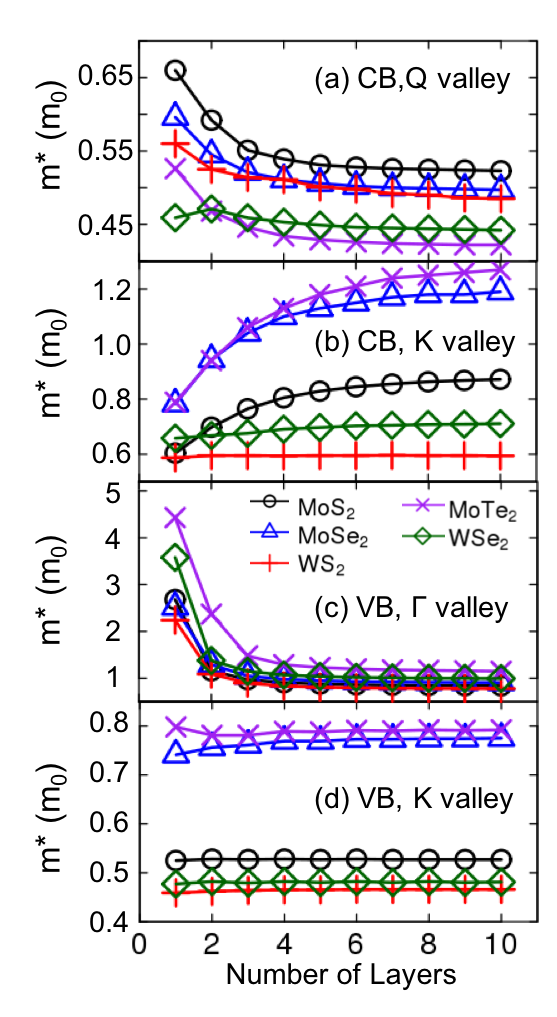}
  \caption{Calculated effective masses along the $\vec{b_1}-2\vec{b_2}$ direction for the conduction band Q (a) and K valley (b), as well as for the valence band $\Gamma$ (c) and K valley (d). }
  \label{fig:eff}
  \end{figure}

  \subsection{Conduction and valence band energies and masses}
  
The conduction band minimum in 2D momentum space for monolayer and 5-layer MoS$_2$ are shown in Figs.~\ref{fig:EK1and5layercontour}. For all TMDs considered in this work, the lowest conduction band hosts valleys at the K points and close to the Q points, respectively. Note that each K-point (Q-point) contributes to 3 (1) Brillouin zones and therefore its valley is twofold (sixfold) degenerate. 
Similarly, all considered TMDs show two valleys in the highest valence band at the $\Gamma$ point and at the K-point~\cite{Yun2012b}.
The relative energies of all these valleys depend on the layer thicknesses, as illustrated in Figs.~\ref{fig:valley} (a) and (b) for conduction and valence bands, respectively.
Most of the TMDs show a transition of the conduction band minimum (valence band maximum) from K to Q (K to $\Gamma$) valley  at around 2 layer thickness~\cite{Zhang2015,Kang2016a}. 
The valley effective masses change with layer thickness as well (see Fig.~\ref{fig:eff}) - very similar to findings of Ref.~\cite{Yun2012b} for MoS$_2$.

  \subsection{Band edge density of states}
  
\begin{figure}[!htbp]
  \centering
  \includegraphics[width=3.4in]{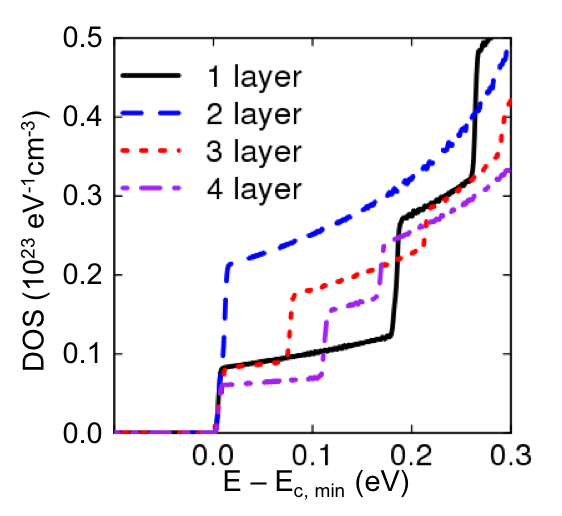}
  \caption{Conduction band density of states in MoS$_2$ layers of various thicknesses.  The 2 layer system has the largest DOS due to the alignment of K and Q conduction band valleys. Each step in the DOS marks an onset of a higher conduction band. The finite slope of the DOS between each step originates from a non-parabolic band dispersion.}
  \label{fig:DOS}
  \end{figure}

For a 2D system, the density of states (DOS) will be proportional to the effective mass. Note that the Q valley conduction band DOS decreases for all TMDs with increasing layer thickness as a direct consequence of the effective mass behavior.
In the valence band, the $\Gamma$ valley effective masses decay with thickness, while the K valley masses stay fairly constant (see Figs.~\ref{fig:eff} (c) and (d)).
The energies of K- and Q-valleys for two layer n-type TMDs are close enough so that both valleys contribute to the DOS at the band edge. 
Once the Q-valley of n-type TMDs (i.e., MoS$_2$~\cite{Li2015}, MoSe$_2~$\cite{Jin2015a}, WS$_2~$\cite{Ovchinnikov2014}) is significantly lower in energy than the K-valley (e.g., for more than 2 layers in the case of MoS$_2$), the DOS at the conduction band edge reduces with increasing layer thickness following the effective mass trend. This is exemplified in Fig.~\ref{fig:DOS} which shows the DOS of MoS$_2$  with varying layer thickness.

      \begin{figure}[!htbp]
  \centering
  \includegraphics[width=3.4in]{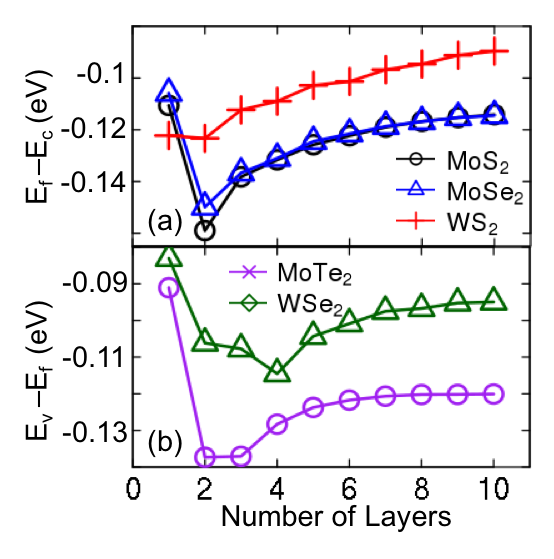}
  \caption{Effective Fermi level $E_f-E_c$ (a) and $E_v-E_f$ (b) of n- and p-type TMD layers with varying thickness and for a given doping density of $\SI{1.5e18}{\per\cubic\centi\metre}$, respectively.}
  \label{fig:efec}
  \end{figure}
  
    \begin{figure}[!htbp]
  \centering
  \includegraphics[width=3.4in]{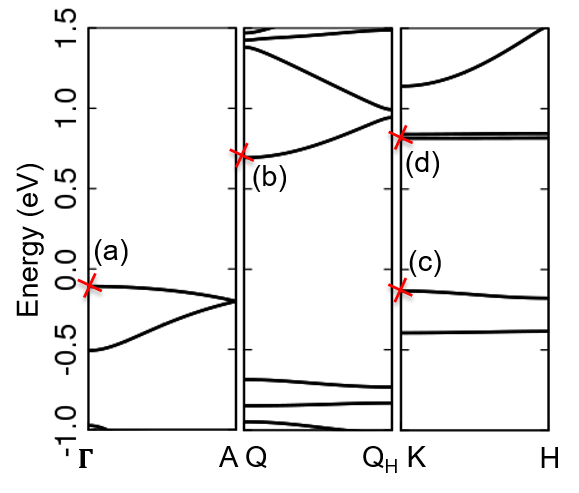}
  \caption{ MoTe$_2$ band diagram along $\vec{b_3}$ starting at the $\Gamma$, Q, and K points defined in Fig.~\ref{fig:EK1and5layercontour}. A, Q$_H$ and H points correspond to $\Gamma$, Q, and K when shifted by $\pi/c\times\vec{b_3}/\lvert b_3\rvert$.} 
  \label{fig:interlayer}
  \end{figure}

    \begin{figure}[!htbp]  \centering
  \includegraphics[width=3.4in]{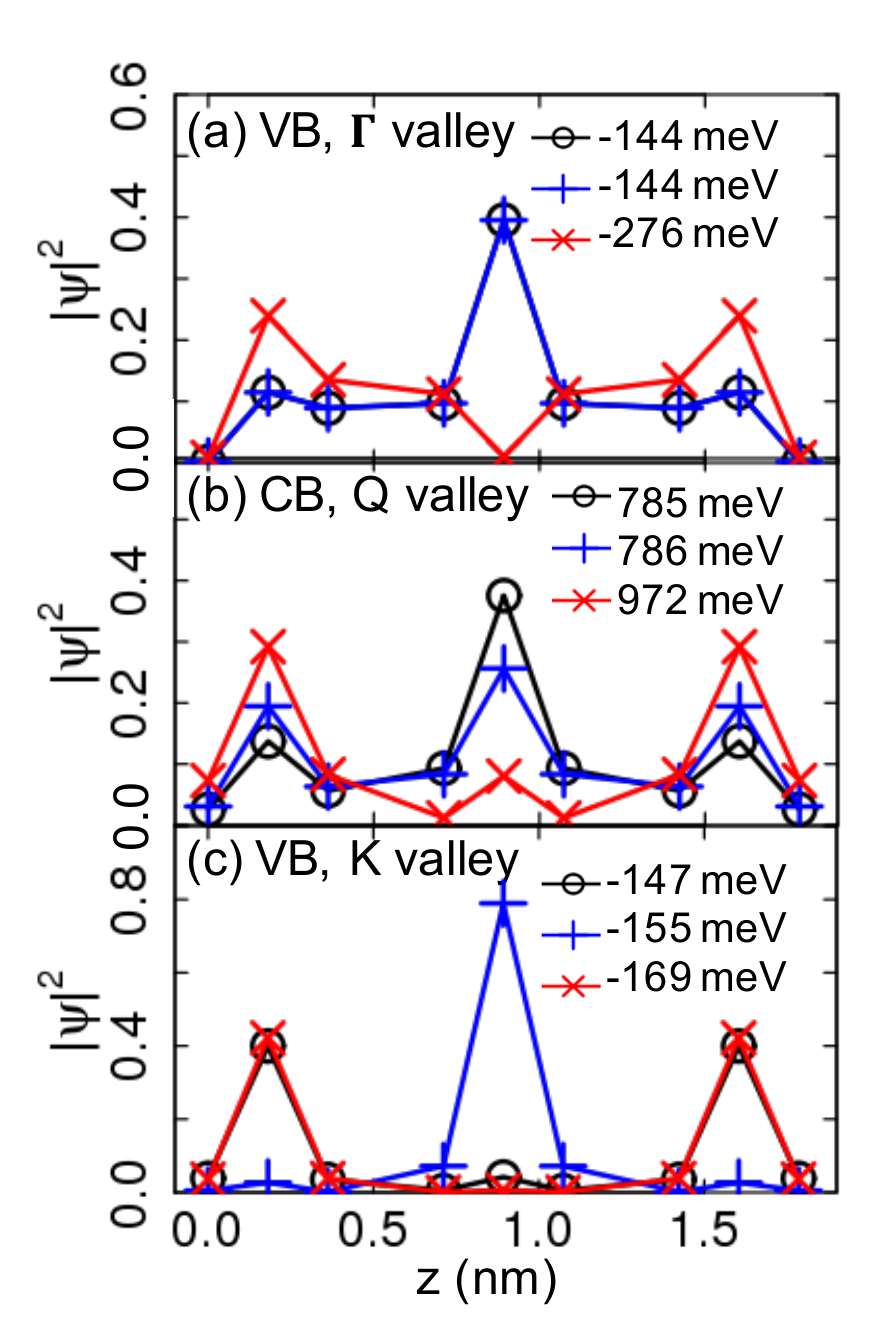}
  \caption{Absolute squared valence band wavefunctions for $\Gamma$ valley of the valence band (a), Q valley of the conduction band (b) and K valley of the valence band (c) of a 3 layer MoTe$_2$ system corresponding to the points (a), (b) and (c) in Fig.~\ref{fig:interlayer}, respectively. The eigenstate energy is used to label the states. Depending on the interlayer and spin orbit coupling strength, the wavefunctions are localized or delocalized. Note that the lines are meant to guide the eye.}
  \label{fig:mote2gammawf}
  \end{figure}

To illustrate the DOS behavior for all considered n-type (p-type) TMDs and thicknesses Fig.~\ref{fig:efec} (a) (Fig.~\ref{fig:efec} (b)) shows the effective Fermi level $E_f-E_c$ $\left(E_v-E_f\right)$ assuming a constant electron (hole) density of \SI{1.5e18}{\per\cubic\centi\metre}. 
Note that the larger the electron (hole) DOS is around the conduction band minimum (valence band maximum), the lower the effective Fermi level has to be to maintain the assumed density. 
All n-type TMDs except WS$_2$ show a maximum DOS at 2 layers thickness when both K- and Q-valleys contribute.
The p-type TMDs show a maximum DOS at 2 and 4 layers for MoTe$_2$ and WSe$_2$, respectively. 
This is the situation when both the $\Gamma$ and the $K$ valleys similarly contribute to the DOS around the Fermi level.

 \subsection{Interlayer hybridization}
 
2H-phase TMDs consist of alternating layers with 2 different orientations of metal-chalcogenide bonds (differing in a 60$^{\circ}$ rotation)~\cite{1969AdPhy..18..193W,Akashi2015}.
If the electrons are subject to a pronounced spin-orbit interaction (e.g., in odd-layer 2H-TMD systems without inversion symmetry~\cite{Akashi2015,Li2013,Komider2013a,Schaibley2016}), electronic states spreading across alternating layers are suppressed~\cite{Jones2014,Gong2013}.
In contrast, the geometrical confinement favors electronic states that are spread across the total device if the interlayer coupling is strong enough~\cite{Xu2014}. 
Figure~\ref{fig:interlayer} shows the bandstructure of an infinite layer MoTe$_2$ system at various symmetry points of the 2D momentum space along $\vec{b_3}$ direction. 
The stronger the interlayer coupling, the stronger curved the respective bands in Fig.~\ref{fig:interlayer} are. 
In Fig.~\ref{fig:interlayer} several points in the bandstructure are labeled. They face different balances between the interlayer coupling strength and the spin-orbit interaction: 
At the valence band $\Gamma$ point (labeled with (a)), the spin-orbit interaction vanishes and only the interlayer coupling determines the shape of the electronic wave functions. 
Here, this coupling is significant and gives an effective mass of -2.31m$_0$ for the top of the valence band. 
Consequently, the highest valence band states resemble typical infinite-barrier quantum well shapes (note they are spin degenerate, see Fig.~\ref{fig:mote2gammawf} (a)). 
At the Q-point of the conduction band (labeled with (b) in Fig.~\ref{fig:interlayer}) the spin-orbit interaction is finite but smaller than the strong interlayer coupling (effective mass m$^{*}$ = 0.53 m$_0$ at Q along $\vec{b_3}$ direction).
Consequently, the shape of the electronic states shown in Fig.~\ref{fig:mote2gammawf} (b) is still comparable with those of Fig.~\ref{fig:mote2gammawf} (a), but the spin degeneracy is lifted. 
The top-valence band states at the K-point (labeled (c) in Fig.~\ref{fig:interlayer}) face similarly strong interlayer coupling (effective mass m$^{*}$ = -1.68$m_0$ at K along K-H direction) and spin-orbit interaction. Consequently, the respective wave functions avoid spreading in alternating layers. Instead, states of equivalent layers (i.e., next-nearest neighbor layers) hybridize into bonding and anti-bonding  states. 
In the case of a 3-layer system, states of only two layers can follow that (depicted in Fig.~\ref{fig:mote2gammawf} (c) with circles for the bonding state and crosses for the antibonding state), while states of the center layer (symbol "+" in  Fig.~\ref{fig:mote2gammawf} (c)) are effectively isolated. In agreement with Ref.~\cite{Akashi2015} the lowest conduction band states at the K-point (labeled with (d) in Fig.~\ref{fig:interlayer}) is found to have a very small interlayer coupling (effective mass m$^{*}$ = -699m$_0$ at K along $\vec{b_3}$ direction), i.e., a coupling smaller than the respective spin-orbit interaction. Therefore, electronic states of individual layers barely interact and are effectively degenerate.
Note that the spin-orbit interaction does not play a significant role in systems with inversion symmetry (i.e., with an even number of layers). Very small spin-orbit coupling effects observed in these systems can be addressed to small p-orbital contributions of chalcogenide atoms to the conduction band~\cite{Komider2013a}. Then, wave functions are exclusively determined by the interlayer coupling strength.
It is worth to emphasize that the wave function effects discussed above are found in all considered TMD materials.

  \begin{figure}[!htbp]
  \centering
  \includegraphics[width=3.4in]{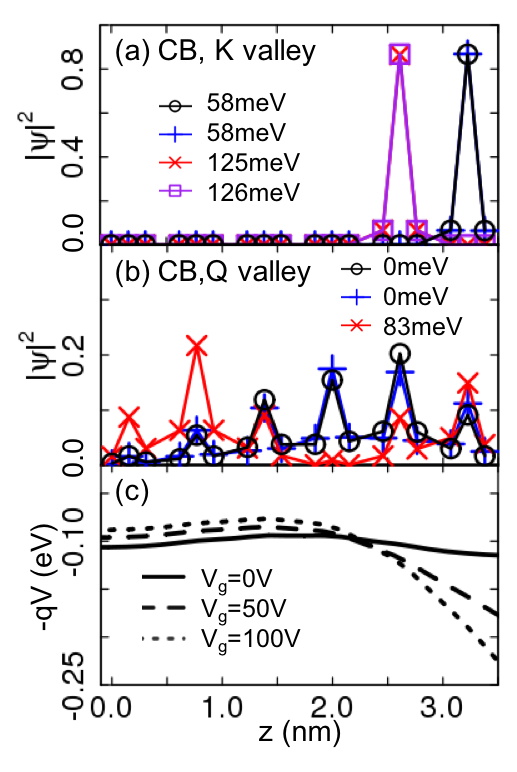}
  \caption{Absolute squared of several conduction band wave functions at the K point (a) and at the Q point (b) for a 6-layer gated MoS$_2$ device depicted in Fig. \ref{fig:structure} at a gate voltage of $V_g$~=~\SI{100}{\volt}. 
The MoS$_2$ system extends from z~=~\SI{0}{\nm} to z~=~\SI{3.6}{\nm}. The potential profiles for various gate voltages are shown in (c). The lines in (a) and (b) are meant to guide the eye.}
  \label{fig:wavefunctionK}
  \end{figure}

  \subsection{Electric gate response}
  
\begin{figure}[!htbp]
  \centering
   \includegraphics[width=3.4in]{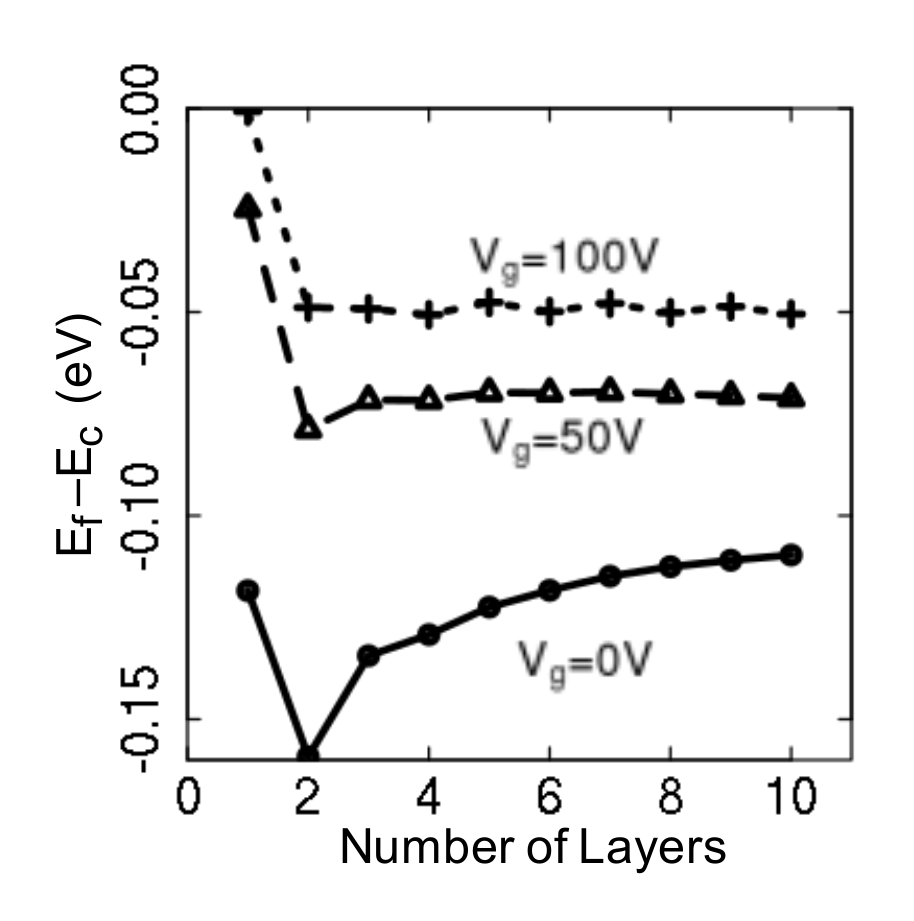}
  \caption{Effective Fermi level of the gated MoS$_2$ layers shown in Fig. \ref{fig:structure} for different thicknesses and applied gate voltages. For this comparison a layer thickness independent doping density of \SI{1.5 e18}{\per\cubic\centi\metre} is assumed. }
  \label{fig:Eclayer}
  \end{figure}

\begin{figure}[!htbp]
  \centering
  \includegraphics[width=3.4in]{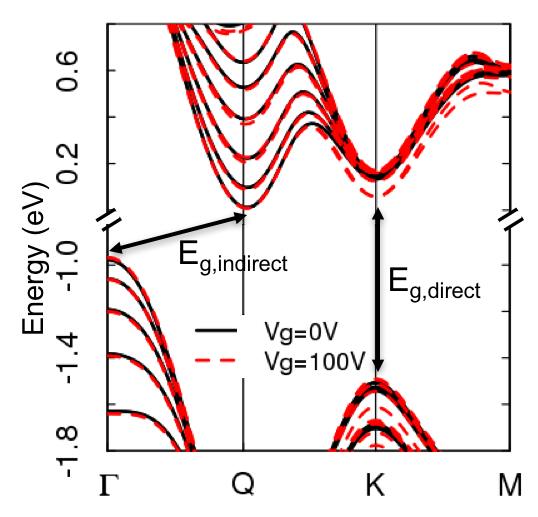}
  \caption{Bandstructures of a 6-layer MoS$_2$ system as shown in Fig.~\ref{fig:structure} for the field free case and when a gate voltage of 100~V is applied. The gate field lifts the K valley degeneracy while the $\Gamma$ and Q valleys remain virtually unaltered. To ease the comparison, the energy offset is chosen to have the Fermi level set to 0 for both voltages.}
  \label{fig:EKfield}
  \end{figure}

The effect of electric gates on TMD layers is exemplified with the absolute squared conduction band wave functions
and the self-consistently solved electrostatic potential of a 6-layer gated MoS$_2$ system shown in Figs.~\ref{fig:wavefunctionK}. 
The energies of K-valley and Q-valley states get closer with the electric field: in the field free case, the bottom of K and Q valleys are separated by more than 100~meV, whereas their energy difference is about 58~meV, as seen in Fig.~\ref{fig:wavefunctionK}. 
Higher gate fields make it energetically more favorable to avoid state delocalization across the total device.
This can be seen for the Q-valley states in Fig.~\ref{fig:wavefunctionK} (b) as their center shifts in the gate field direction. 
The electrostatic potential profiles for several different gate voltages are shown in Fig.~\ref{fig:wavefunctionK}~(c). 
In these and all other considered cases of this work, the gate field is screened within about 1~nm penetration depth. Consequently, the thinner the TMD system is, the larger is its response to the applied gate field. 
This is exemplified in Fig.~\ref{fig:Eclayer} for the effective Fermi level as a function of gate bias and layer thickness. 
Note that the monolayer results of this figure still assume completely screened gate field in the vacuum, in spite of the pronounced penetration depth. 
Thus the monolayer results are given for the sake of completeness only and to ease comparison with the effective Fermi levels shown in Fig.~\ref{fig:efec}.
For higher gate fields and TMDs thicker than the field penetration length, the gate induced shift of effective Fermi level becomes independent of the layer thickness (see Fig.~\ref{fig:Eclayer}). 
Since K-valley states are localized within monolayers, they face a layer-dependent effective electric field. 
Accordingly, the K-valley degeneracy gets lifted by electric gates, as illustrated in Fig.~\ref{fig:wavefunctionK}~(a) and Fig.~\ref{fig:EKfield}. 
Note that the Q-valley conduction band energies and $\Gamma$ valley valence band energies remain virtually unaffected by the electric gate (see Fig.~\ref{fig:EKfield}).

   \begin{figure}[!htbp]
   \centering
   \includegraphics[width=3.4 in]{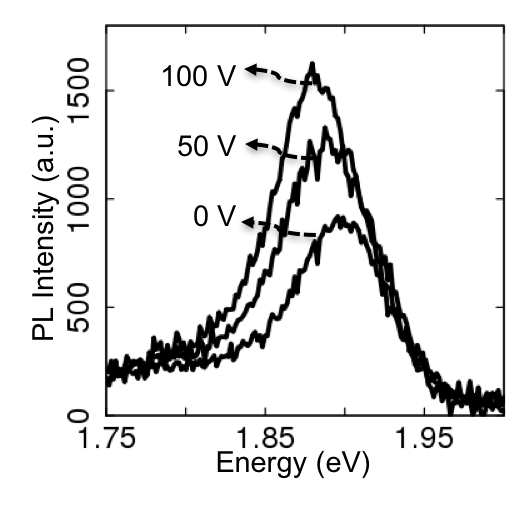}
  \caption{
  Experimental photoluminescence (PL) spectra (equivalent to "peak A" in Ref.~\cite{Mak2010}) of a gated 6 layer MoS$_{2}$ structure for various gate voltages. The increase of the PL amplitude with the gate voltage qualitatively agrees with band structure changes predicted in Fig.~\ref{fig:EKfield} as discussed in the main text.}
   \label{fig:AFMPLexp}
   \end{figure}

  \begin{figure}[!htbp]
  \centering
  \includegraphics[width=3.4in]{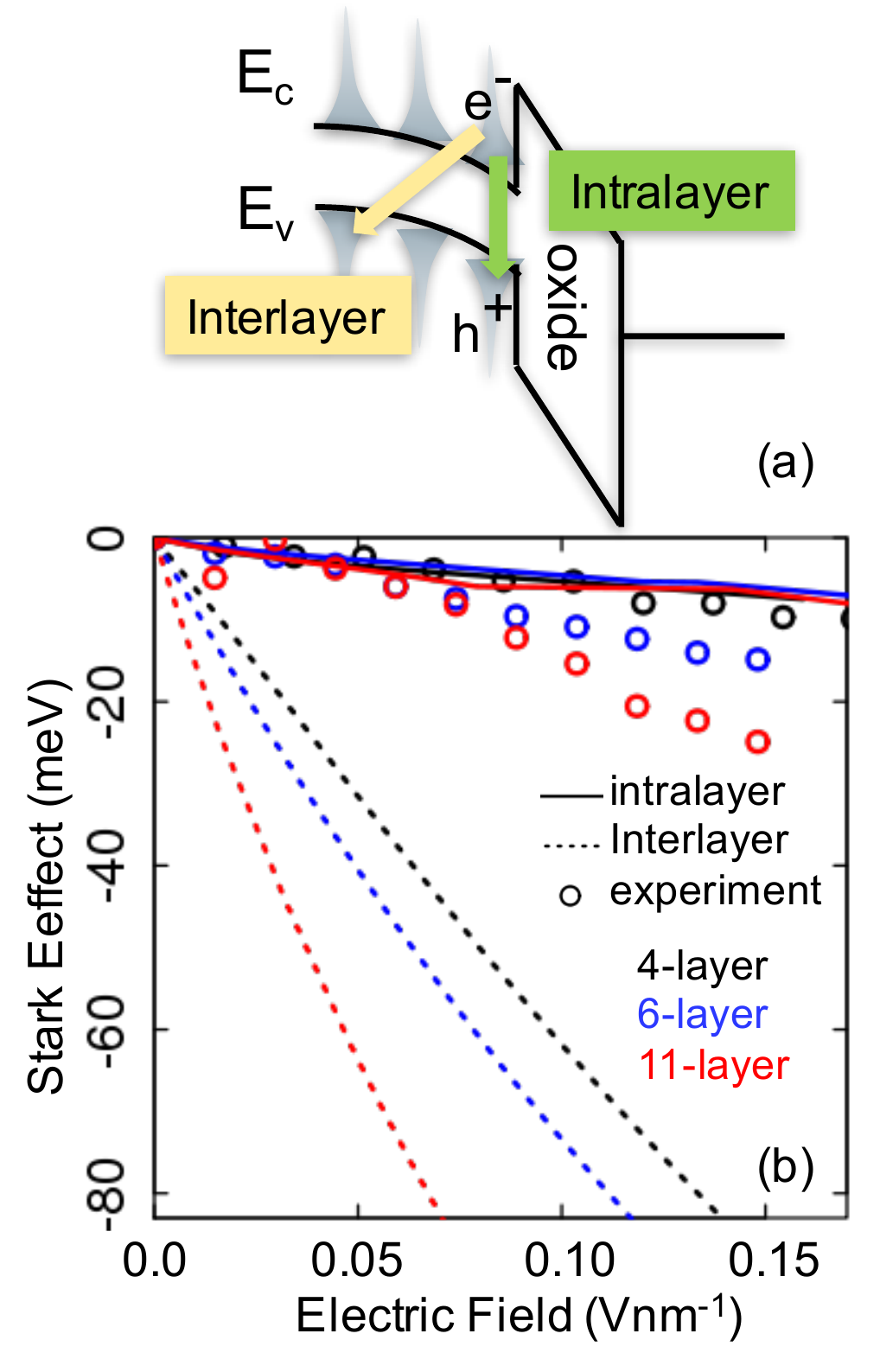}
  \caption{ (a) Schematic of the interlayer and intralayer exciton recombinations. (b) The relative change of experimentally observed direct band gap exciton energies (symbols) as a function of applied gate bias for various MoS$_2$ systems agrees well with the theoretically predicted energy differences of conduction and valence band K-valley states of identical layers (solid lines). In contrast, the calculated conduction and valence K-valley energy differences of maximally separated layers (dotted) significantly exceed the experimentally observed Stark effects.}
  \label{fig:PLexp}
  \end{figure}

  \subsection{Quantum confined Stark effect}

Figure~\ref{fig:EKfield} also shows that the quantum confined Stark effect of the K-valley states reduces the direct band gap at the K-point. Similar effects were observed for direct band gap excitons in Ref.~\cite{Klein2016,Kim2016} as well as in the experiments of this work: the photoluminesence peaks of direct band gap excitons show a red-shift with increasing gate voltage (see Fig.~\ref{fig:AFMPLexp}).
Figure~\ref{fig:AFMPLexp} also shows an increase of the PL amplitude with increasing gate bias. This qualitatively agrees with the bandstructure results of Fig.~\ref{fig:EKfield}: The gate field barely changes the K-valley valence band edge, but it lowers the K-valley energy of the conduction band. This results in an increase of the K-valley electron density with the gate bias, while the K-valley hole density is approximately constant. Since the photoluminescence amplitude is approximately proportional to the product of electron and hole K-valley density (see e.g. Ref.~\cite{Piprek2010,PhysRev.87.387}), it increases with the gate bias. 

It had been discussed in literature (Ref.~\cite{Chu2015,Klein2016,Chen2016}) whether the direct band gap excitons are recombining within individual layers or across different layers (illustrated in Fig.~\ref{fig:PLexp} (a)). 
To clarify the nature of the excitons and shed more light on this question, Fig.~\ref{fig:PLexp} (b) compares the field induced changes of the experimentally observed exciton energies with NEMO5 results. 
Since the NEMO5 calculations do not include exciton binding energies, differences of K-valley conduction and valence band states of the same layer and of maximally separated layers are used to represent intralayer and interlayer excitons, respectively.
For comparability of experimental and NEMO5 results, the transition energy changes in Fig.~\ref{fig:PLexp} (b) are shown relative to the field free case. 
For both, the interlayer and intralayer transitions NEMO5 predicts a finite Stark effect, but only the intralayer transition Stark effects agree qualitatively with the experimental data.
Note that NEMO5 calculations of the optical matrix elements~\cite{chuang1995physics} (not in the figure) showed two orders of magnitude higher probability for intralayer transitions then for interlayer ones. 
It is also worth to mention that our experiments did not show any Stark effect for indirect band gap excitons - in agreement with the theoretical results in Fig.~\ref{fig:EKfield} that show virtually gate field independent Q-valley and $\Gamma$-valley energies.
In summary, these results suggest that MoS$_2$ excitons preferably perform intralayer transitions.

\section{Conclusion}
Electronic wave functions and bandstructures in 2H-TMD structures were analyzed in the MLWF representation of the nanodevice simulation tool NEMO5.
Hybridization of electronic states across multiple layers was shown to depend on the balance of spin-orbit coupling and interlayer coupling strength.
This balance varies strongly with the electronic momentum.
Conduction band K-valley states are found to be confined in individual monolayers.
In contrast, valence band K-valley states are delocalized in equivalent layers for systems with finite spin-orbit coupling or across the total device when the spin-orbit coupling disappears. 
This K-state hybridization can be lifted with electric gate fields.
The design range of the spin-orbit interaction, the interlayer coupling, the effective Fermi levels and effective masses are carefully assessed.
Experimental data of the quantum confined Stark effect of direct band gap, interlayer and intralayer exciton photoluminesence were reproduced with NEMO5.
Intralayer excitons were identified as the major source for photoluminesence signals - in agreement with a previous study Ref.~\cite{Klein2016}.

\section*{Acknowledgment}
The work is supported by NSF EFRI-1433510. We also acknowledge the Rosen Center for Advanced Computing at Purdue University for the use of their computing resources and technical support. This research is part of the Blue Waters sustained-petascale computing project, which is supported by the National Science Foundation (award number ACI 1238993) and the state of Illinois. Blue Waters is a joint effort of the University of Illinois at Urbana-Champaign and its National Center for Supercomputing Applications. This work is also part of the “Accelerating Nano-scale Transistor Innovation with NEMO5 on Blue Waters” PRAC allocation support by the National Science Foundation (award number OCI-0832623). 
This work was partially supported by the National Science Foundation’s MRSEC program (DMR-1121262) and made use of its Shared Facilities at the Materials Research Center of Northwestern University. This work made use of the EPIC facility of Northwestern University’s NU\textit{ANCE} Center and the NUFAB facility, which have received support
from the Soft and Hybrid Nanotechnology Experimental (SHyNE) Resource (NSF ECCS-1542205); the MRSEC program (NSF DMR-1121262) at the Materials Research Center; the International Institute for Nanotechnology (IIN); the Keck Foundation; and the State of Illinois. N.P.S. acknowledges support as an Alfred P. Sloan Research Fellow.
J.M. acknowledges support from NSERC.

 \bibliographystyle{apsrev4-1k}
 \bibliography{library.bib}

\end{document}